\pdfoutput=1

\documentclass[11pt]{article}

\usepackage[final]{coling}

\usepackage{times}
\usepackage{latexsym}
\usepackage{subfig}
\usepackage{longtable}
\usepackage[T1]{fontenc}

\usepackage[utf8]{inputenc}

\usepackage{microtype}

\usepackage{inconsolata}

\usepackage{graphicx}
\usepackage{multirow}
\usepackage{mathtools}
\usepackage[normalem]{ulem}
\useunder{\uline}{\ul}{}
\title{KG-Retriever: Efficient Knowledge Indexing for Retrieval-Augmented Large Language Models}

\author{
 \textbf{Weijie Chen\textsuperscript{1}},
 \textbf{Ting Bai\textsuperscript{1}\thanks{Corresponding author.}},
 \textbf{Jinbo Su\textsuperscript{2}},
 \textbf{Jian Luan\textsuperscript{3}},
 \textbf{Wei Liu\textsuperscript{3}},
 \textbf{Chuan Shi\textsuperscript{1}}
\\
\\
 \textsuperscript{1}Beijing University of Posts and Telecommunications,
 \\
 \textsuperscript{2}University of Science and Technology Beijing,
 \\
 \textsuperscript{3}Xiaomi Corporation.
}
\begin{document}
\maketitle
\begin{abstract}


Large language models with retrieval-augmented generation 
encounter a pivotal challenge in intricate retrieval tasks, e.g., multi-hop question answering, which requires the model to navigate across multiple documents and generate comprehensive responses based on fragmented information. 
To tackle this challenge, we introduce a novel Knowledge Graph-based RAG framework with a hierarchical knowledge retriever, 
termed KG-Retriever. 
The retrieval indexing in KG-Retriever is constructed on a hierarchical index graph that consists of a knowledge graph layer and a collaborative document layer.
The associative nature of graph structures is fully utilized to strengthen intra-document and inter-document connectivity, 
thereby fundamentally alleviating the information fragmentation problem and meanwhile improving the retrieval efficiency in cross-document retrieval of LLMs.
With the coarse-grained collaborative information from neighboring documents and concise information from the knowledge graph, KG-Retriever achieves marked improvements on five public QA datasets, showing the effectiveness and efficiency of our proposed RAG framework.


\end{abstract}

\section{Introduction}

Large Language Models (LLMs) with Retrieval-Augmented Generation (RAG) have achieved initial success in generating accurate responses, especially in knowledge-intensive tasks. By using a retrieval component to incorporate relevant information from a vast corpus of external documents, the RAG technique has become the mainstream way to alleviate hallucination issues in the response generation of LLMs~\cite{yang2023empower,ding2023harnessing}.
Nonetheless,
when engaging in intricate retrieval tasks, e.g., multi-hop question answering, the model encounters significant difficulties in extracting pertinent answers solely from a single document. 
It requires the model to navigate across multiple documents and generate comprehensive responses according to the fragmented information from multiple relevant documents. 
For example, in the multi-hop question
“What are the trend and major factors contributing to dry eye syndrome in children, and what preventive measures can be taken?”, most existing RAG-based LLMs inevitably suffer from incomplete retrieval knowledge due to the disability of reasoning over different documents. 
Recent RAG studies~\cite{feng2024retrieval,shao2023enhancing} attempt to disassemble the retrieval process into iterative retrieval steps by using the generated content from the last iteration as the query to retrieve relevant documents in the next round. 
Such approaches improve the inferential capabilities of LLMs and enhance the retrieval quality to some extent, but they still face the challenge of the escalating computational costs caused by multiple iterative retrieval steps. Besides, due to the discreteness of each iteration, such methods may still face poor retrieval performance in integrating information across different documents.


To enhance retrieval quality while maintaining RAG's efficiency, we propose a novel knowledge graph-based RAG framework with a hierarchical knowledge retriever, termed \textbf{KG-Retriever}. 
Specifically, KG-Retriever is built based on a Hierarchical Index Graph (\textbf{HIG}) that consists of a knowledge graph layer and a collaborative document layer (as shown in Fig.~\ref{fig: overview}). In the knowledge graph layer, entities and relations in a document are extracted by LLMs, which enhances the internal information structuring of individual documents. 
In the collaborative document layer, the document-level graph establishes connections based on their semantic similarity, improving the cross-document knowledge correlations. 
The entity-level and document-level information in HIG enable our KG-Retriever to leverage the associative nature of graph structures to strengthen intra-document and inter-document connectivity, thereby fundamentally augmenting the retrieval efficiency of LLMs.


Based on the HIG, the retrieval process starts at the document layer and subsequently broadens to include information from neighboring documents. The indirect relevant information sourced from neighboring documents offers supplementary validation against potential false relevancy between queries and documents.
Then, entity-level matching is conducted within the KG layer of candidate documents. The triplets in the KG layer of candidate documents further provide concise information to mitigate data noise and reinforce the logical coherence of retrieval outcomes through inter-entity linkages. The collaborative information in both the document layer and KG layer in KG-Retriever work together to provide more comprehensive information and exclude irrelevant information through two rounds of matching, leading to a marked improvement in the quality and credibility of generated content. Finally, the selected triplets in KG, combined with the original query, are fed into the large language models to generate the final responses. The contributions of our work are as follows:


\begin{itemize}
\item We construct a novel retrieval indexing based on a Hierarchical Index Graph (HIG), which establishes intra-document and inter-document connectivity on the document layer and KG layer, enhancing the internal information structuring of individual documents and cross-document knowledge correlations.
 
\item We propose a hierarchical knowledge retriever (KG-Retriever) to facilitate retrieval-augmented LLMs. KG-Retriever is capable of leveraging the supplementary validation information from neighborhoods of different documents and reinforcing the logical coherence through inter-entity linkages, achieving marked improvement in the quality and credibility of generated content of LLMs.

\item We conduct extensive experiments on five representative open-domain QA datasets.
Compared with RAG approaches with single or multi-iteration retrieval steps, our method achieves the SOTA model performance under a single retrieval setting and attains substantial advancements in terms of efficiency.
\end{itemize}

\section{Related Work}
In this section, we introduce the concepts related to our study briefly, including the naive RAG, the advanced RAG, and the graph-based RAG in the research area of LLMs.

\subsection{Retrieval-Augmented Generation}
Retrieval Augmented Generation (RAG) methodologies address the issue of hallucinations of large language models (LLMs) in knowledge-intensive tasks by supplementing them with relevant external information~\cite{ding2024survey,gao2023retrieval}. This technique is segmented into three phases: indexing, retrieval, and generation. Naive RAG approaches~\cite{lewis2020retrieval,guu2020retrieval}, as an archetype, initiates by splitting text into chunks and embedding each chunk into vector representations for indexing. Subsequently, in the retrieval stage, it matches the query’s embedding to these vectorized chunks based on semantic similarity. In the final generation phase, the retrieved chunks, combined with the original query, serve as contextual input for LLMs to generate a refined response. Despite marked enhancements, it may lose advantages in complex multi-hop question-answering scenarios, as simple indexing and retrieval strategies inadequately integrate cross-document information, resulting in diminished retrieval accuracy and substandard response quality.

\subsection{Advanced RAG}

Recent studies have attempted to enhance retrieval quality by fine-tuning retrieval models with task-specific data~\cite{shao2023enhancing} or by discriminating among retrieved contents through end-to-end fine-tuning of LLMs~\cite{asai2023self}. However, such methods are difficult to apply in zero-shot settings, as the highly tailored fine-tuning process limits their ability to extend to unseen tasks or data. 
Efforts have also been made to enhance the interplay between retrieval and generation phases. For instance, ITRG~\cite{feng2024retrieval} employs an iterative strategy where each iteration's output is used in the subsequent retrievals, enhancing both the generated content's quality and retrieval precision. 
IRCOT~\cite{trivedi2022interleaving} integrates retrieval with the Chain-of-Thought (CoT) approach, guiding retrieval by CoT and refining CoT with retrieved content. Despite their contributions, these iterative retrieval-enhanced strategies suffer substantial computational costs due to the repetitive retrieval and are limited by the disjoint nature of information gathered across iterations.

Different from these approaches, our methodology focuses on the indexing phase and proposes retrieval on a hierarchical index graph. It leverages graph connectivity to achieve coherent, in-depth retrieval within a single retrieval step, thus avoiding the computational costs of iterative searches. 

\subsection{Graph-based RAG}
The integration of graphs with LLMs and the RAG technique has recently elicited significant attention, with plenty of directions being proposed~\cite{peng2024graph}. These include using LLMs for knowledge graph creation~\cite{trajanoska2023enhancing,edge2024local}, completion~\cite{yao2023exploring}, and knowledge editing~cite{shi2024retrieval}. 
The study Graph RAG~\cite{he2024g,kang2023knowledge} proposed a Graph RAG method to retrieve subgraphs for the field of Graph Question Answering (GraphQA). 
Recent studies have proposed similar concepts for enhancing the overall performance of RAG systems by creating graph structures. 
For instance, KGP~\cite{wang2024knowledge} creates a knowledge graph over multiple documents with nodes symbolizing passages or document structures. Some researchers~\cite{edge2024local} have explored building graph-based texts to improve the performance of LLMs in Query-Focused Summarization. 

Different from the above methods, our work extracts entity relationships from all documents to construct an entity-level knowledge graph. Based on this knowledge graph, cross-document associations are constructed, forming a hierarchical graph structure. We propose a novel single-retrieval RAG framework with different retrieval strategies on the hierarchical index graph, achieving both the efficiency and accuracy of the RAG framework.

\begin{figure*}
\centering
\includegraphics[width=\linewidth]{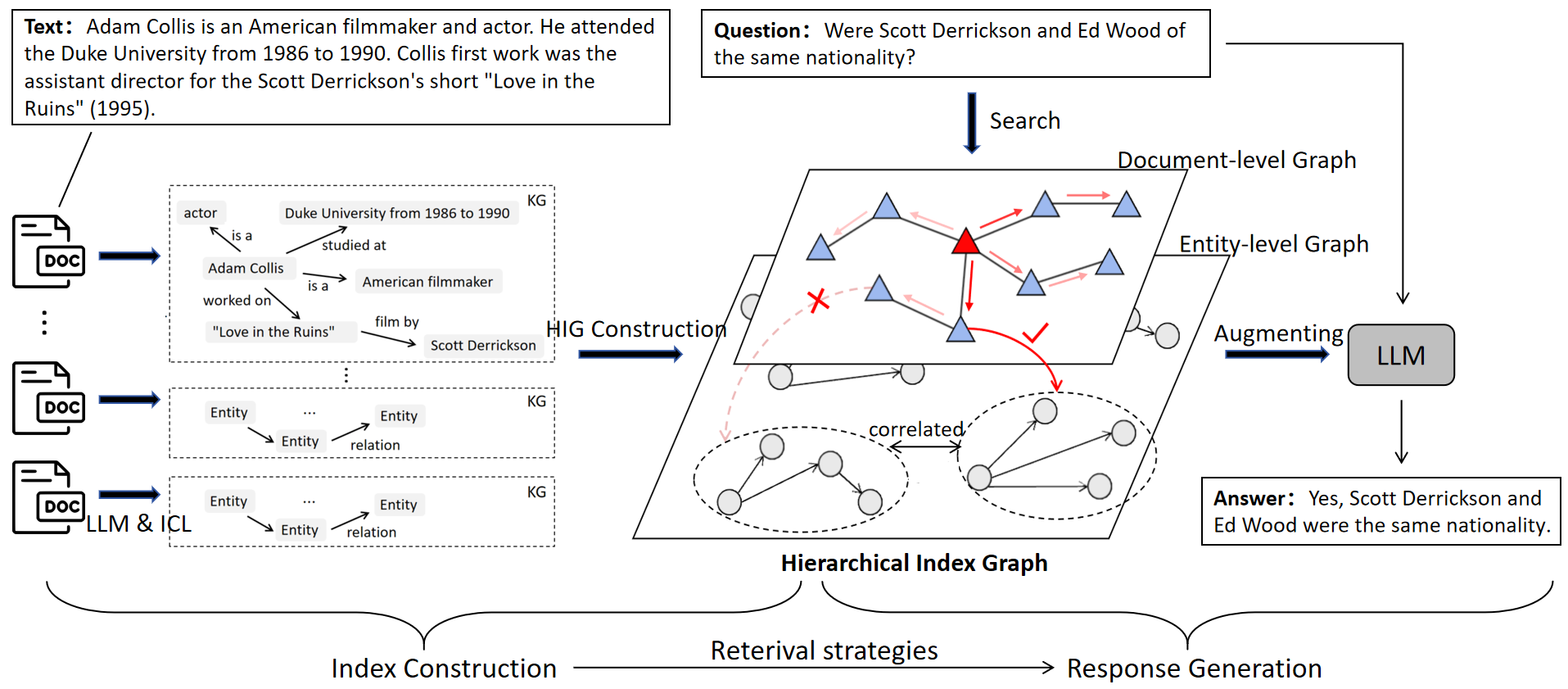}
  \caption{The overview architecture of KG-Retriever. It consists of three components: the indexing construction component based on a hierarchical index graph (HIG), the knowledge retrieval component, and the response generation component.}
  \label{fig: overview} 
\end{figure*}

\section{Methodology}

The following sections introduce the overview architecture of KG-Retriever and the details of each component's implementation.

\subsection{Overview Architecture}

As shown in Figure~\ref{fig: overview}, KG-Retriever consists of three components, i.e., the indexing construction component, the knowledge retrieval component, and the response generation component. 
The retrieval index is constructed on a Hierarchical Index Graph (HIG), which consists of a document graph and an entity graph. 
Both internal structure within documents and the broader interconnections across documents can be retrieved on HIG. 
In the knowledge retrieval component, two rounds of matching at the document layer and KG layer work together to provide more comprehensive information and exclude irrelevant information.
Specifically, 
different retrieval strategies are used to obtain the relevant information in associated documents within the document-level graph. 
Subsequently, an entity-level matching process within these identified documents meticulously extracts KG triples that are closely related to the query. 
In the response generation component, the retrieved knowledge information from the KG, combined with the original question, serves as inputs to feed into the LLMs, generating precise responses.


\subsection{KG-Retriever}

The information retrieval process in KG-Retriever is conducted with a Hierarchical  Index Graph (HIG), in which the associative nature of graph structures is used to strengthen intra-document and inter-document connectivity. The construction of HIG and retrieval strategies are introduced in the following sections. 

\subsubsection{HIG Construction}

To enhance the coherence of intra-document information and inter-document informational relationships, we design a novel retrieval index structure based on a Hierarchical Index Graph (HIG). 
HIG consists of a Knowledge Graph (KG) layer and a collaborative document layer. 
The connection in the document layer improves knowledge correlations across documents.
The information in the KG layer further enhances text comprehension and facilitates intricate intra-document information interactions.

\noindent\textbf{Entity-Level KG Construction}.
The entities in the documents are organized on a KG, in which the precise representation of entities and relationships enhances semantic comprehension and facilitates the exploitation of KG information.  
Considering that recent advancements in LLMs, such as the closed-source GPT-4~\cite{achiam2023gpt} and open-source alternatives such as Llama~\cite{touvron2023llama} and Qwen~\cite{bai2023qwen}, show remarkable capabilities in understanding complex language tasks, including entity identification and relationship inference from unstructured text, we use the in-context learning paradigm and design prompts to enhance LLMs' ability to generate accurate knowledge graphs of documents. We choose Qwen-72B to extract the KG in our experiments. Examples of the prompt templates to construct KG are shown in Table~3 and Table~4 in the Appendix.

\noindent\textbf{Documents-level Graph Construction}.
The above entity-level KG models relationships within a document. To enhance retrieval efficiency and improve RAG's capability of integrating information across multiple documents, we construct a document-level graph upon entity-level knowledge graphs.
Specifically, we employ a pre-trained Language Model (LM), e.g., Roberta-large\footnote{https://huggingface.co/FacebookAI/roberta-large} 
to enpublic documents. The semantic representation of a document $d$ is denoted as $\textbf{v}_d$.
The edge in the document graph is established according to the similarity between documents. The edges between document $d$ and its neighbors are identified by:
\begin{equation}
\underset{j \in\{1,2, \ldots, M\}}{\operatorname{argmax_{K}}} (\operatorname{CosSim}(\textbf{v}_d,\textbf{v}_j)),
\label{eq:topk}
\end{equation}
where $\textbf{v}_d$ and $\textbf{v}_j$ denote the semantic representations of the document $d$ and $j$, respectively. $M$ is the number of documents. $K$ is a hyper-parameter that selects the $K$ most similar neighbors according to the cosine similarity of two document vectors. The edges are established between document $d$ and the Top-K similar neighbors.


\subsubsection{Retrieval Strategies} 
To achieve comprehensive information extraction that integrates both the macroscopic perspective of documents and the microscopic details of entities, we proposed three retrieval strategies built upon HIG that contain two stages: document-level retrieval and KG-level retrieval. 

\noindent\textbf{Document-level Retrieval}.
The matching process between the query and documents is based on their semantic similarity. 
We use the same pre-trained language model, e.g., Roberta-large, to encode the semantics of queries. For a query, the retrieval is conducted to identify the Top-N documents that are most similar to the semantic vector of the query.
To address potential issues of spurious relevance inherent in direct semantic similarity matching, we collect the neighboring documents of the identified Top-N documents on the document graph into our final candidate set. While these neighboring documents may not be directly relevant to the query, their proximity to the highly pertinent Top-N documents suggests they may contain information indirectly related or supplementary to the query. This approach enriches the contextual understanding of the query, thereby enhancing the comprehensiveness and depth of the generated responses. Three distinct document-level collaboration strategies are proposed to enhance the retrieval process:

\begin{itemize}
    \item \textbf{One-Hop Collaboration}:
        This strategy integrates neighboring documents into the candidate set without discrimination, encompassing all one-hop neighbors to ensure broad applicability.
        
    \item \textbf{Attentive Collaboration}: This strategy includes one-hop neighbors while incorporating an attentive filtering mechanism (controlled by $\lambda$) to refine selections of triples in the KG layer and meanwhile enhance the relevance of the retrieved knowledge.  Attention weights are computed as the cosine similarity of the two documents' semantic vectors.
    
    \item \textbf{Multi-Hop Collaboration}: Going beyond one-hop neighbors, this strategy incorporates multi-hop neighbors and utilizes attention-based weights, i.e., multiply attention weights in different hops, to improve precision and coverage in retrieval tasks. 
\end{itemize}

The three strategies enable our framework to be robust and adaptable to meet the different retrieval needs of tasks. The One-Hop Collaboration strategy has the fastest processing time and is suitable for most scenarios. The Attentive Collaboration strategy, by introducing the attention mechanism, better filters information and improves retrieval precision. The Multi-Hop Collaboration strategy, by expanding the receptive field, is capable of handling more complex tasks.

\noindent\textbf{KG-level Retrieval}.
After obtaining the related documents through the three collaboration strategies, we further match the entity nodes connected to these candidate documents. 
For an entity $e$ in candidate documents, its related triples in KG are retrieved to generate the answer, denoted as:
\begin{equation}
\text{Retrieved?} = 
\begin{cases}
yes & \text {if} \  w*\operatorname{CosSim}(\textbf{v}_e,\textbf{v}_q) > \lambda; \\
no & \text {else},
\end{cases}
\label{eq:lambda}
\end{equation}
where $\textbf{v}_e$ and $\textbf{v}_q$ are the semantic representations of the entity $e$ and the query $q$ respectively. $w$ is the attention weight of the three collaboration strategies in the document collaboration process. In One-Hop Collaboration, $w$ is set to 1, while in the attentive methods, $w$ is the attention weights computed by the semantic similarity of documents. $\lambda$ is the threshold value to filter KG information, ensuring that only entities with attentive similarity exceeding this value are considered.
We set the maximum of retrieved triples as $T$ for efficiency consideration.

After the document-level collaboration and KG-level collaboration,  the retrieved triples in KG combine with the original query, serving as inputs to the LLMs for generating the response.

\section{Experiments}

\subsection{Experimental Settings}
\textbf{Datasets}.
To verify the effectiveness of KG-Retriever in intricate QA tasks,
five open-domain question-answering datasets are used in our experiments, including HotpotQA~\cite{yang2018hotpotqa}, MuSiQue~\cite{trivedi-etal-2022-musique}, 2WikiMultilHopQA~\cite{ho-etal-2020-constructing}, CRUD-QA1, and CRUD-QA2~\cite{lyu2024crud}. 
The evaluations on these datasets cover multiple perspectives: multilingual capacity (covering both Chinese and English), diverse question scenarios (multi-hop QA and single-hop QA tasks), and a range of response types (including both short-form and long-form generation tasks).

\begin{itemize}
    \item \textbf{HotpotQA}: The HotpotQA is a multi-hop English QA dataset, with a collection of related Wikipedia context and question-answer pairs. The answers in this dataset are short-form texts and the related Wikipedia contexts contain both positive relevant and negative relevant passages. 

    \item \textbf{MuSiQue}: The MuSiQue is designed to support the development and evaluation of models that perform multiple steps of reasoning to answer a question, with a collection of related context and question-answer pairs. 
    
    \item \textbf{2WikiMultilHopQA}: The 2WikiMultilHopQA is a multi-hop English QA dataset, with a collection of related Wikipedia context and question-answer pairs. The related Wikipedia contexts contain both positive and negative relevant passages. 

    \item\textbf{CRUD-QA1 \& CRUD-QA2}: The CRUD datasets were constructed by crawling the latest high-quality news data from the mainstream news websites in China. Specifically, CRUD-QA1 is a single-hop QA dataset, while CRUD-QA2 is a multi-hop QA dataset. Both datasets contain a collection of related news context and question-answer pairs, where the answer is long-form texts.
\end{itemize}

\noindent\textbf{Evaluation Metrics}.
For HotpotQA, MuSiQue and 2WikiMultilHopQA datasets, we follow the settings in~\cite{yang2018hotpotqa}: use the collection of related context for each pair mixed up as the retrieval corpus and adopt Exact Match (EM) as the evaluation metric. 
For CRUD-QA1 and CRUD-QA2 datasets, we follow the settings in~\cite{lyu2024crud}: use the collection of related news context mixed together as the retrieval corpus and adopt BLEU and Rouge-L for evaluations.

\noindent\textbf{Baselines}. We compare KG-Retriever with representative LLMs and RAG methods, including:

\begin{itemize}
    \item \textbf{Naive LLM}~\cite{bai2023qwen}: It prompts
    LLMs to directly generate the final answer without retrieval.
    \item \textbf{LLM with CoT}: It prompts LLMs to generate both the chain-of-thought (CoT) reasoning process and the final answer.
    \item \textbf{Graph-guided reasoning}~\cite{park2023graph}: 
     It proposes a graph-guided CoT prompting method that guides the LLMs to the correct answer with graph verification steps.
     \item \textbf{BM25}: It uses the question as a query to retrieve N paragraphs by the traditional sparse retrieval method BM25. 
    \item \textbf{DenseRetriever}: It uses the question as a query to retrieve N paragraphs by the dense retrieval method. For English QA scenarios, we used $\emph{Roberta-large}$ as the retriever, while for Chinese QA scenarios, we used $\emph{bge-base}$ as the retriever. 
    \item \textbf{ITRG}~\cite{feng2024retrieval}: It employs an iterative strategy where each cycle’s output informs subsequent retrievals, enhancing both the generated content’s quality and retrieval precision. 
    \item \textbf{ITER-RETGEN}~\cite{shao2023enhancing}: It follows an iterative strategy, merging retrieved and generated contents in each iteration as a whole, thereby overcoming the shortcomings of traditional iterative strategy in information integration. 
    \item \textbf{KGP}~\cite{wang2024knowledge}: It segments documents into passages and uses KNN to build a graph. Then, it introduces an LLM as an agent to perform retrieval on the graph, determining the nodes to visit in the next iteration.
\end{itemize}

The above methods cover different kinds of RAG approaches in
LLMs research:  naive method directly generating answers without retrieval (Naive LLMs), using chain-of-thought reasoning (LLMs with CoT), and employing graph-guide CoT prompting (Graph-guided reasoning). Traditional retrieval methods like BM25 and dense retrieval methods are also discussed. Additionally, some advanced iterative strategies such as ITRG, ITER-RETGEN, and KGP are also compared. 
\textbf{Note that} we conduct our experiments on the zero-shot settings. For fair comparisons, the RAG methods~\cite{asai2023self,he2024g} that need Supervised Fine-Tuning (SFT) are not compared in our experiments. 

\noindent\textbf{Implementation Details}.
For each baseline method, a grid search is applied to find the optimal settings. 
We report the result of each method with its optimal hyperparameter settings.
For hyperparameters in our model, the number of neighbors $K$ in the document-level graph construction (see Eq.~\ref{eq:topk}), the number of retrieved documents during Document-level Collaboration $N$, the max retrieved entity count $T$ in KG-level Collaboration, and the retrieval threshold $\lambda$ for entity retrieval (see Eq.~\ref{eq:lambda}) and has been set as follows: 
\{K=2, N=3, T=20, $\lambda$=0.1\} for HotpotQA,
\{K=3, N=3, T=30, $\lambda$=0.1\} for MuSiQue,
\{K=3, N=3, T=30, $\lambda$=0.1\} for 2WikiMultilHopQA,
\{K=1, N=3, T=10, $\lambda$=0.4\} for CRUD-QA1,
\{K=2, N=3, T=15, $\lambda$=0.3\} for CRUD-QA2. 
The code is publicly available on https://github.com/BAI-LAB/KG-Retriever.

\begin{table*}[]
\centering
\caption{Performance comparison on five datasets. The metric 'Time' represents the average running time for generating the response in LLMs.}
\resizebox{\linewidth}{!}{
\begin{tabular}{c|cc|cc|cc|ccc|ccc}
\hline \hline
\multirow{2}{*}{Datasets} & \multicolumn{2}{c|}{HotpotQA}   & \multicolumn{2}{c|}{MuSiQue}    & \multicolumn{2}{c|}{2WikiMultiHopQA} & \multicolumn{3}{c|}{CRUD-QA1}                    & \multicolumn{3}{c}{CRUD-QA2}                     \\ \cline{2-13} 
                          & EM             & Time           & EM             & Time           & EM                & Time             & BLEU           & Rouge-L        & Time           & BLEU           & Rouge-L        & Time           \\ \hline \hline
Naive LLM                 & 0.102          & 0.69s          & 0.040          & 0.31s          & 0.170             & 0.28s            & 0.073          & 0.24           & 1.75s          & 0.079          & 0.237          & 0.95s          \\
COT + LLM                 & 0.172          & 2.11s          & 0.060          & 1.32s          & 0.180             & 1.45s            & 0.035          & 0.146          & 2.11s          & 0.030          & 0.151          & 4.06s          \\
Graph-guided reasoning    & 0.197          & 40.03s         & 0.070          & 42.30s         & 0.210             & 39.08s           & 0.311          & 0.393          & 34.59s         & 0.095          & 0.244          & 27.18s         \\ \hline
BM25                      & 0.236          & 1.25s          & 0.070          & 0.67s          & 0.250             & 0.73s            & 0.209          & 0.509          & 1.08s          & 0.069          & 0.233          & 1.37s          \\
DenseRetriever            & 0.282          & 0.25s          & 0.050          & 0.35s          & 0.220             & 0.45s            & 0.23           & 0.395          & 1.06s          & 0.155          & 0.259          & 1.44s          \\
ITRG (5-Iteration)        & 0.306          & 10.99s         & 0.120          & 9.60s          & \uline{0.320}             & \uline{6.30s}            & \uline{0.333}          & \uline{0.457}          & \uline{14.7s}          & \uline{0.172}          & \uline{0.277}          & \uline{13.55s}         \\
ITER-RETGEN (3-Iteration) & \uline{0.323}          & \uline{6.65s}          & \uline{0.170}          & \uline{9.71s}          & 0.290             & 10.03s           & 0.24           & 0.551          & 6.65s          & 0.069          & 0.236          & 8.95s          \\
KGP (3-Iteration)         & 0.278          & 6.34s          & 0.140          & 6.98s          & 0.220             & 7.03s            & 0.275          & 0.376          & 7.07s          & 0.155          & 0.235          & 7.30s          \\ \hline 
\textbf{KG-Retriever}            & \textbf{0.328} & \textbf{0.93s} & \textbf{0.210} & \textbf{1.21s} & \textbf{0.350}    & \textbf{0.74s}   & \textbf{0.449} & \textbf{0.611} & \textbf{0.95s} & \textbf{0.233} & \textbf{0.353} & \textbf{1.46s} \\
KG-Retriever (Attention)  & 0.322          & 1.14s          & 0.200          & 0.88s          & 0.340             & 0.83s            & 0.458          & 0.600          & 0.99s          & 0.239          & 0.357          & 1.63s          \\
KG-Retriever (Multi-Hop)  & 0.328          & 1.19s          & 0.210          & 1.03s          & 0.350             & 0.83s            & 0.458          & 0.600          & 1.15s          & 0.238          & 0.354          & 2.20s          \\ \hline 
\end{tabular}
}
\end{table*}

\subsection{Main Results}
We conduct experiments on five datasets to show the effectiveness of KG-Retriever in QA tasks.
The experiments are conducted on the open-sourced multilingual large language model $\emph{Qwen1.5-7b}$. As shown in Table 1, we can see that:

(1) The LLMs with RAG significantly outperform the methods that do not leverage retrieval augmentation. LLMs incorporate techniques such as Chain-of-Thought (CoT) or graph reasoning, failing to break the inherent limitations of LLMs in knowledge-intensive QA tasks.

(2) For the LLMs with the RAG technique, the advanced RAG methods like ITRG, ITER-RETGEN, and KGP perform better than the naive single-retrieval methods (e.g., BM25 and DenseRetriever), showing the necessary of using iterative retrieval steps to obtain supplementary information from multiple documents.

(3) Our KG-Retrieval achieves state-of-the-art performance compared with all the baselines on five datasets.  With only one retrieval step, our KG-Retriever shows better model performance than the methods that retrieve multiple times.
Besides, our KG-Retriever shows a great advantage in generation efficiency: it is 6 to 15 times faster than these iterative methods, i.e., ITRG (11.6 times) and ITER-RETGEN (8.4 times), in generating the response for a question.

(4)  In KG-Retrieval with three retrieval strategies, the attentive collaboration strategy performs better on CURD-QA1 and CURD-QA2 datasets due to the fact that KG information in the more similar documents is selected to provide precise supplementary information. Besides, the 2-hop collaboration makes slight improvements over the 1-hop collaboration, showing that our hierarchical index graph could provide adequate information for LLMs, ensuring high retrieval efficiency.





\subsection{Experimental Analysis}
We conduct in-depth analyses of our method to quantitatively demonstrate its effectiveness, 
including ablation studies, the performance on different LLM backbones, and hyperparameter analyses. 

\subsubsection{Ablation Study}
We conduct ablation studies of our framework to identify the key factors in KG-Retriever. 
The variants of our framework are: 
\begin{itemize}
\item w/o Entity Graph: it solely constructs the document-level graph with inter-document collaborative information.
\item w/o Document Graph: only entity-level knowledge graph is used to enhance the retrieval results. 
\end{itemize}

The experimental results are shown in Table.~\ref{fig: abs}, we can see that (1) modifications to the hierarchical graph structure led to a huge degradation of model performance, showing that both intra-document and inter-document information learned in our hierarchical index graph are useful in enhancing overall performance;
(2) the degradation of the variant without document-level information is much larger than without entity information in HotpotQA, MuSiQue, and CRUD-QA1 datasets. While in CRUD-QA2, the conclusion is on the contrary. The reason may lie in that for the simple answer (short-form text response) in HotpotQA and CRUD-QA1 datasets, the collaboration from similar documents is more important than concrete entities, while for the complex answer (long-form text response) in CRUD-QA2 dataset, incorporating more specific knowledge is more helpful to enhance the quality of retrieval results.
This indicates the necessity to conduct inter-document and intra-document information retrieval of KG-Retriever.

\begin{table}[]
\caption{The performance comparison of the degeneration variants of our KG-Retriever.}
\resizebox{\linewidth}{!}{%
\begin{tabular}{c|c|c|c|c}
\hline \hline 
Datasets                  & Metrics & w/o Entity & w/o Document & Our   \\ \hline \hline
HotpotQA                  & EM      & 0.282      & 0.160        & 0.328  \\ 
MuSiQue                   & EM      & 0.180      & 0.140        & 0.210  \\ 
2WikiQA                   & EM      & 0.330      & 0.320        & 0.350  \\ 
\multirow{2}{*}{CRUD-1}   & BLEU    & 0.427      & 0.275        & 0.449  \\
                          & Rouge-L & 0.587      & 0.326        & 0.611  \\ 
\multirow{2}{*}{CRUD-2}   & BLEU    & 0.103      & 0.209        & 0.233  \\
                          & Rouge-L & 0.192      & 0.296        & 0.353  \\ \hline
\end{tabular}%
}
\label{fig: abs} 
\end{table}

\begin{figure*}[h]
    \subfloat[HotpotQA]{\includegraphics[width=0.25\linewidth]{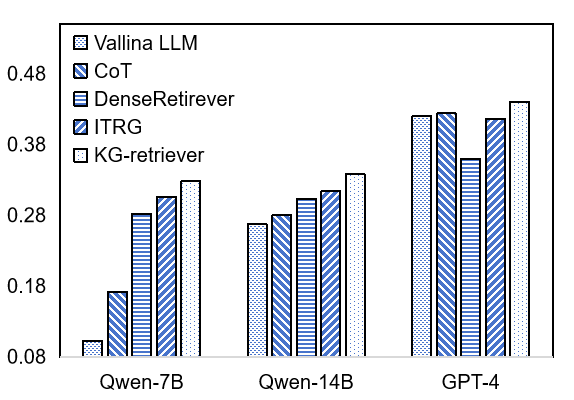}}
    \subfloat[MuSiQue]{\includegraphics[width=0.25\linewidth]{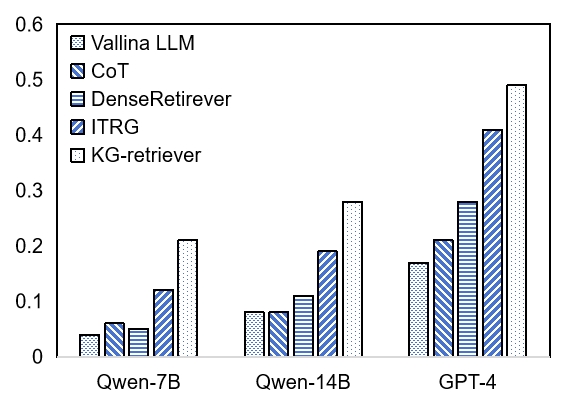}}
    \subfloat[2WikiMultiHopQA]{\includegraphics[width=0.25\linewidth]{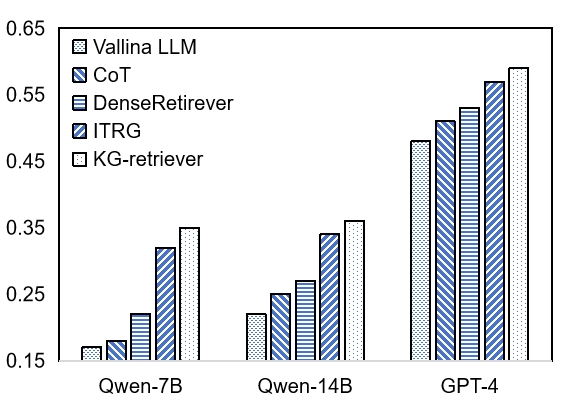}}
    \subfloat[CRUD-QA2]{\includegraphics[width=0.25\linewidth]{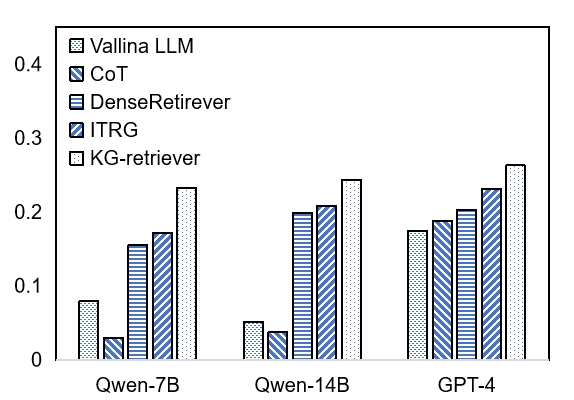}}
    \caption{The performance of RAG methods with different LLM backbones (Qwen-7B, Qwen-14B and GPT-4).}
    \label{fg:backbone}
\end{figure*}

\begin{figure*}[h]
    \subfloat[Parameter $K$]{\includegraphics[width=0.33\linewidth]{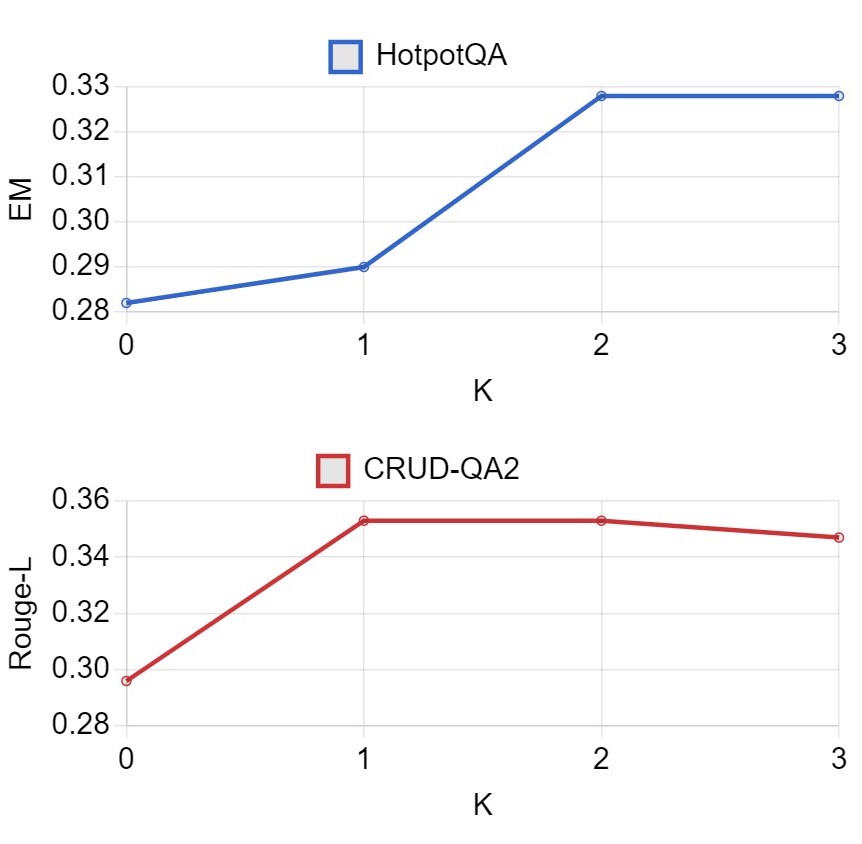}}
    \subfloat[Parameter $T$]{\includegraphics[width=0.33\linewidth]{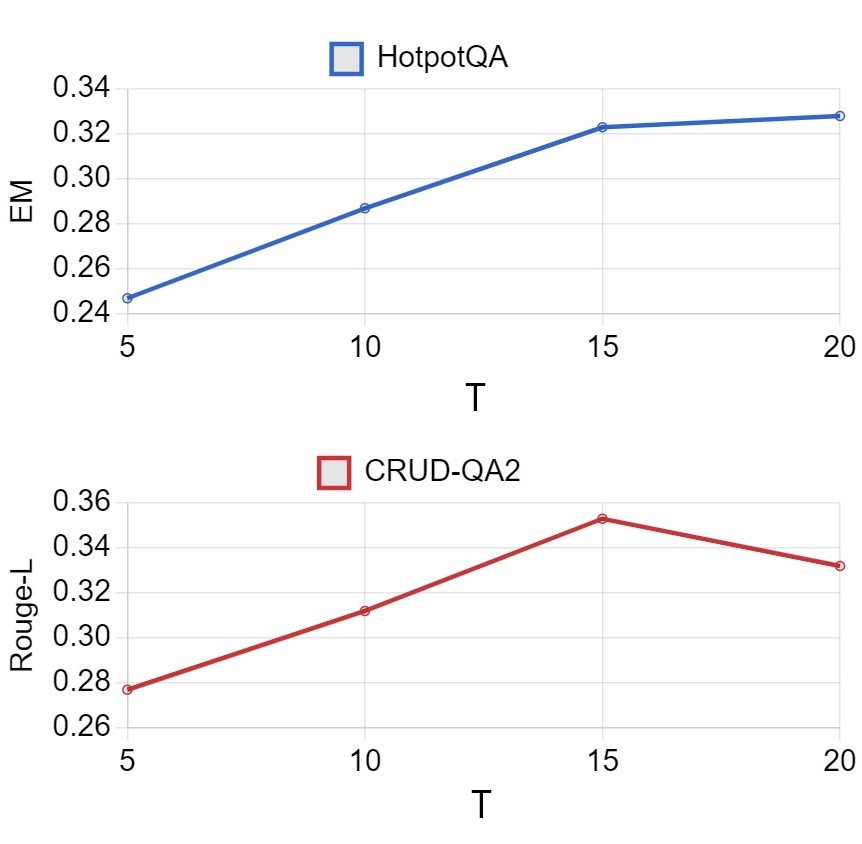}}
    \subfloat[Parameter $\lambda$]{\includegraphics[width=0.33\linewidth]{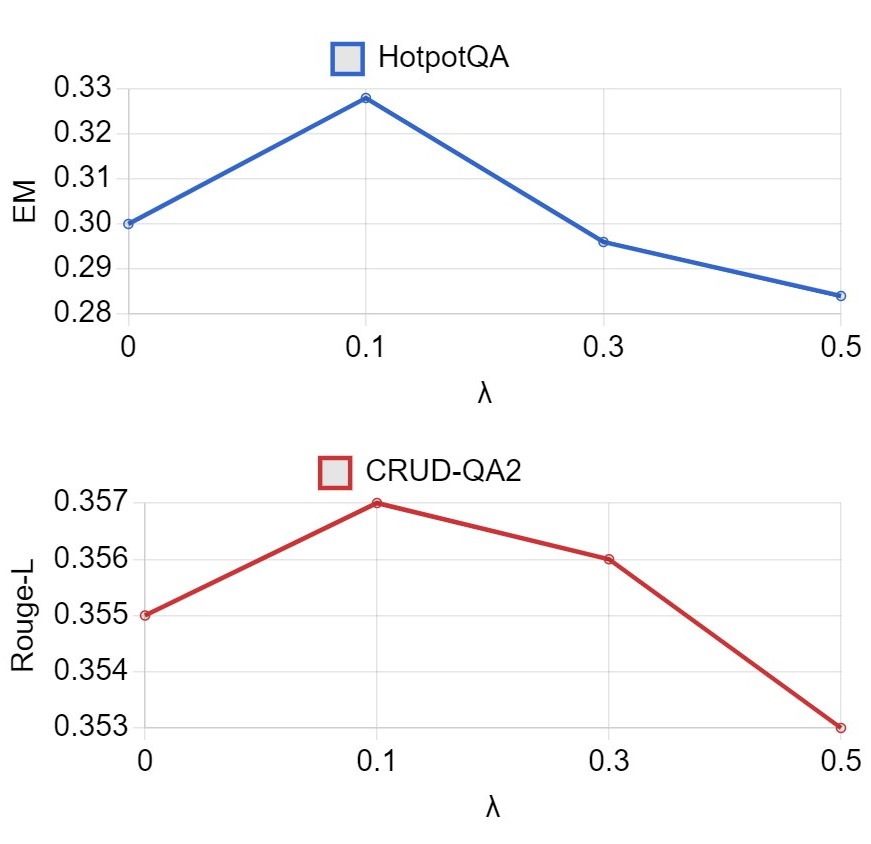}}
    \caption{Hyperparameter Analysis on HotpotQA and CRUD-QA2 datasets. $K$ is a hyper-parameter that selects the most similar neighbors, and $T$ and $\lambda$ are the hyper-parameter to control the number of retrieved triples.}
    \label{fg:Stablility}
\end{figure*}

\subsubsection{Impacts of LLM Backbones}
LLMs equipped with more parameters show greater capability to deal with complex tasks. Considering the potential influence of different LLM backbones on the model performance, we analyze the performance of our RAG framework on more powerful LLM backbone models, i.e., the Qwen-14b and GPT-4. 
As illustrated in Fig~\ref{fg:backbone}, we can observe that:
(1) Compared with the experimental results on Qwen-7B and Qwen-14B, GPT-4 achieves significant improvements on all tasks, showing its greater ability in natural language understanding.
(2) Even on the powerful large language model (GPT-4), our KG-based RAG framework (KG-Retriever) makes further improvements over the baseline methods on all datasets, consistently demonstrating its effectiveness.


\subsubsection{Hyperparameters Analysis}
Our framework leverages hyperparameters to influence the retrieved information, including the number of collaborative documents $K$ in Eq.~\ref{eq:topk}, the maximum of retrieved triples $T$ in Section 3.2, and the similarity threshold $\lambda$ in Eq.~\ref{eq:lambda} for entity retrieval. In this section, we delve into analyses of the impacts of these parameters on the overall performance of our framework. The results are shown in Fig.~\ref{fg:Stablility}, we can see that:
(1) In subfigure (a), incorporating more documents improves the performance of our framework. After reaching a saturation point, the increment of neighboring documents may incorporate noise and result in degradations of model performance. 
(2) In subfigure (b), increasing the maximum number of retrieved triplets can provide richer contextual information and improve model performance. However, too many triplets may also introduce irrelevant information, negatively affecting the model performance.
(3) In subfigure (c), adjusting the threshold for entity retrieval serves a dual purpose: it filters out irrelevant information, leading to performance enhancements. However, excessively stringent thresholds can inadvertently discard valuable information, thereby causing a decline in performance.

\section{Conclusion}
In this paper, we introduce a novel retrieval-augmented generation framework, KG-Retriever. 
We construct an effective retrieval indexing on the the hierarchical index graph (HIG) and carefully design the corresponding retrieval strategies on it.
Indexing from HIG, KG-Retriever is capable of capturing both intra-document and inter-document information, making it easily adaptive to retrieving precisely collaborative information in different QA tasks.
Besides, KG-Retriever provides a way to obtain abundant knowledge in once retrieval process, largely improving the generation efficiency of LLMs.

\section{Limitations}
While KG-Retriever requires lower reasoning demands on the backbone model compared to Interactive RAG, it still necessitates that the backbone model possesses sufficient inferential capability to derive relevant information from the knowledge graph. 
Besides, although our indexing methodology enhances retrieval and inference efficiency, constructing the index itself requires computational resources. However, since this process can be completed offline, it does not affect real-time inference in real applications. 
In addition to QA tasks, we will make comprehensive model validation in other NLP domains, e.g., fact verification and commonsense reasoning, to verify the effectiveness of our RAG framework.
Additionally, static indexing structures may not be optimal for dynamic corpus, future work will explore dynamic indexing structures to address this limitation. 


\bibliography{coling_latex}

\appendix
\label{sec:appendix}

\noindent  
\begin{table*}[h]
\centering
\begin{tabular}{p{\textwidth}}
\caption{Prompt template to extract triples in KG on HotpotQA dataset.} \\
\hline
\textbf{Instruction}: 
You are an NLP assistant. Given a piece of text, you need to analyze its semantic information and generate a knowledge graph. Your output consists only of triples, considering only the text content, and avoiding newline characters. For example, (entity; relationship; entity),(entity; relationship; entity). The knowledge graph should be comprehensive, covering all information in the text. \\
\# \\
\textbf{Text}:  
Adam Collis is an American filmmaker and actor. He attended the Duke University from 1986 to 1990 and the University of California, Los Angeles from 2007 to 2010. He also studied cinema at the University of Southern California from 1991 to 1997. Collis first work was the assistant director for the Scott Derrickson's short "Love in the Ruins" (1995). In 1998, he played "Crankshaft" in Eric Koyanagi's "Hundred Percent". \\
\textbf{Triples:}
(Adam Collis; nationality; American),(Adam Collis; profession; filmmaker),(Adam Collis; profession; actor),(Adam Collis; education; Duke University),(Adam Collis; education; University of California, Los Angeles),(Adam Collis; education; University of Southern California),(Adam Collis; attended; Duke University),(Adam Collis; attended; University of California, Los Angeles),(Adam Collis; attended; University of Southern California),(Adam Collis; first work; assistant director),(Adam Collis; work; "Love in the Ruins"),("Love in the Ruins"; director; Scott Derrickson),(Adam Collis; work date; 1995),(Adam Collis; role; "Crankshaft"),("Hundred Percent"; director; Eric Koyanagi),(Adam Collis; work; "Hundred Percent"),(Adam Collis; work date; 1998) \\
\# \\
\textbf{Text}:
Tyler Bates (born June 5, 1965) is an American musician, music producer, and composer for films, television, and video games. Much of his work is in the action and horror film genres, with films like "Dawn of the Dead, 300, Sucker Punch," and "John Wick." He has collaborated with directors like Zack Snyder, Rob Zombie, Neil Marshall, William Friedkin, Scott Derrickson, and James Gunn. With Gunn, he has scored every one of the director's films; including "Guardians of the Galaxy", which became one of the highest grossing domestic movies of 2014, and its 2017 sequel. In addition, he is also the lead guitarist of the American rock band Marilyn Manson, and produced its albums "The Pale Emperor" and "Heaven Upside Down". \\
\textbf{Triples}:
(Tyler Bates; birthdate; June 5, 1965),(Tyler Bates; nationality; American),(Tyler Bates; profession; musician),(Tyler Bates; profession; music producer),(Tyler Bates; profession; composer),(Tyler Bates; works in; films),(Tyler Bates; works in; television),(Tyler Bates; works in; video games),(Tyler Bates; specializes in; action and horror film genres),(Tyler Bates; notable films; "Dawn of the Dead"),(Tyler Bates; notable films; "300"),(Tyler Bates; notable films; "Sucker Punch"),(Tyler Bates; notable films; "John Wick"),(Tyler Bates; collaborations; Zack Snyder),(Tyler Bates; collaborations; Rob Zombie),(Tyler Bates; collaborations; Neil Marshall),(Tyler Bates; collaborations; William Friedkin),(Tyler Bates; collaborations; Scott Derrickson),(Tyler Bates; collaborations; James Gunn),(Tyler Bates; collaborations; James Gunn),(Tyler Bates; collaborations; James Gunn),(Tyler Bates; scored; "Guardians of the Galaxy"),("Guardians of the Galaxy"; release year; 2014),("Guardians of the Galaxy"; grossing; high),("Guardians of the Galaxy"; sequel; released in 2017),(Tyler Bates; lead guitarist; Marilyn Manson),(Marilyn Manson; music albums; "The Pale Emperor"),(Marilyn Manson; music albums; "Heaven Upside Down") \\
\hline
\end{tabular}
\label{tb:p_hp}
\end{table*}

\noindent  
\begin{table*}[h]
\begin{tabular}{p{\textwidth}}
\caption{Prompt template to extract triples in KG on CRUD-QA1 and CRUD-QA2 datasets.} \\
\hline
\textbf{Instruction:}
You are an NLP assistant. Given a piece of text, you need to analyze its semantic information and generate a knowledge graph. Your output consists only of triples, considering only the text content, and avoiding newline characters. For example, (entity; relationship; entity),(entity; relationship; entity). The knowledge graph should be comprehensive, covering all information in the text. \\
\#\\
\textbf{Text}:
To foster a positive atmosphere regarding children's eye health across society and continually advance comprehensive efforts to prevent and control myopia among children and adolescents, the National Health Commission has decided to launch the nationwide "Bright Vision Initiative" – a health promotion campaign for preventing and controlling myopia in children and adolescents. They have also issued the "Ten Core Knowledge Points for Preventing and Controlling Myopia in Children and Adolescents."The theme of this initiative is "Prioritize Children's Eye Health, Safeguard Children's Clear Vision." Emphasizing prevention as the primary approach, it aims to shift the focus to earlier stages, advocating for joint actions by families and society as a whole. The goal is to create a visual-friendly environment that promotes eye care and protection, ensuring children have bright futures.

\textbf{Triples}: 
(The National Health Commission; launches; "Enlightenment Action" - health promotion activities for preventing and controlling myopia in children and adolescents), ("Enlightenment Action" - health promotion activities for preventing and controlling myopia in children and adolescents; theme; emphasizing children's eye care; guarding children's clear vision "sight" field), (The National Health Commission; issues; "Top Ten Core Knowledge on Preventing and Controlling Myopia in Children and Adolescents"), ("Top Ten Core Knowledge on Preventing and Controlling Myopia in Children and Adolescents"; type; scientific knowledge on preventing and controlling myopia), (The National Health Commission; requires; conducting social publicity and health education),(Social publicity and health education; utilizing; internet, radio and television, newspapers and magazines, posters and bulletin boards, training seminars), (Health education; objective; popularizing scientific knowledge on preventing myopia), (Health education; target; the general public), (Health education; focus; "Top Ten Core Knowledge on Preventing and Controlling Myopia in Children and Adolescents"), (Health education; method; innovative educational approaches and media), (Health education; purpose; enhancing specificity, precision, and effectiveness), (Health education; tool; internet media) \\
\#\\
\textbf{Text:} 
Pushing forward nationwide fitness is a long-term task that requires persistent efforts. Sustaining and stimulating people's enthusiasm for fitness, meeting diverse demands for sports consumption, all require meticulous efforts. In recent years, various regions have taken multiple measures to encourage public participation in sports and fitness, making considerable efforts in this regard. To boost enthusiasm for exercise among the public, some places have come up with innovative ideas. Recently, Xi'an, Shaanxi Province, allocated 5 million yuan in sports electronic consumption vouchers. Citizens who receive these vouchers can use them at 173 sports venues across the city. The issuance of sports consumption vouchers has not only encouraged public participation in fitness activities but also boosted the operation of sports venues, further unleashing the potential for sports consumption. The goal is not just to encourage regular fitness but also to ensure that people know how to exercise effectively.
\\
\textbf{Triples}: 
(National Fitness; nature; long-term task), (National Fitness; requires; persistent efforts), (Stimulating fitness enthusiasm; goal; meeting diverse sports consumption demands), (Stimulating fitness enthusiasm; method; meticulous efforts), (Local actions; initiative; taking multiple measures to encourage public participation in sports and fitness), (Local actions; provision; providing diverse sports services), (Local actions; effort; meticulous efforts), (Xi'an, Shaanxi Province; action; distributing 5 million yuan worth of electronic sports consumption vouchers),
 \\
\hline
\end{tabular}
\label{tb:p_crud}
\end{table*}

\end{document}